\documentclass[pra,twocolumn]{revtex4}
\usepackage{epsfig}
\usepackage{graphicx}
\usepackage{amsmath}
\usepackage{natbib}
\newcommand{\bn}{\begin{eqnarray}}
\newcommand{\en}{\end{eqnarray}}
\newcommand{\eml}{\end{multline}}
\newcommand{\bml}{\begin{multline}}

\begin{document}

\title {Quasi-continuous variable quantum computation with collective spins in multi-path interferometers}
 \author{Tom\'{a}\v{s} Opatrn\'{y}}
 \affiliation{Optics Department, Faculty of Science, Palack\'{y} University, 17. Listopadu 12,
 77146 Olomouc, Czech Republic}

\date{\today }
\begin{abstract}
Collective spins of large atomic samples trapped inside optical resonators can carry quantum information that can be processed in a way similar to quantum computation with continuous variables. It is shown here that by combining the resonators in multi-path interferometers one can realize coupling between different samples,  and that polynomial Hamiltonians can be constructed by repeated spin rotations and twisting induced by dispersive interaction of the atoms with light. Application can be expected in efficient simulation of quantum systems.
\end{abstract}

\maketitle

%%%%%%%%%%%%%%%%%  F I G U R E %%%%%%%%%%%%%%%%%%%%%%
\begin{figure}%[h!]
\centerline{\epsfig{file=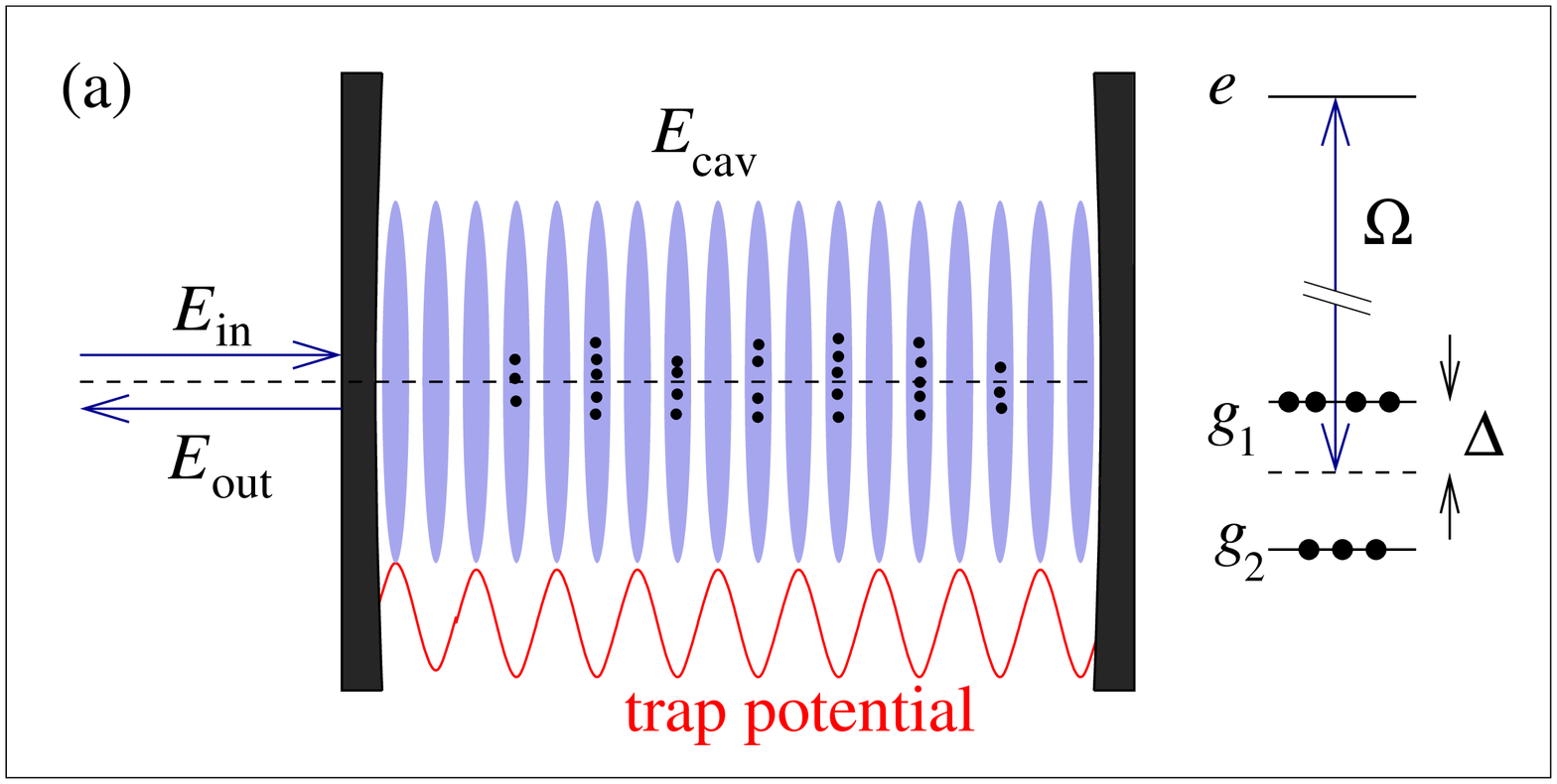,width=0.9\linewidth}}
\centerline{\epsfig{file=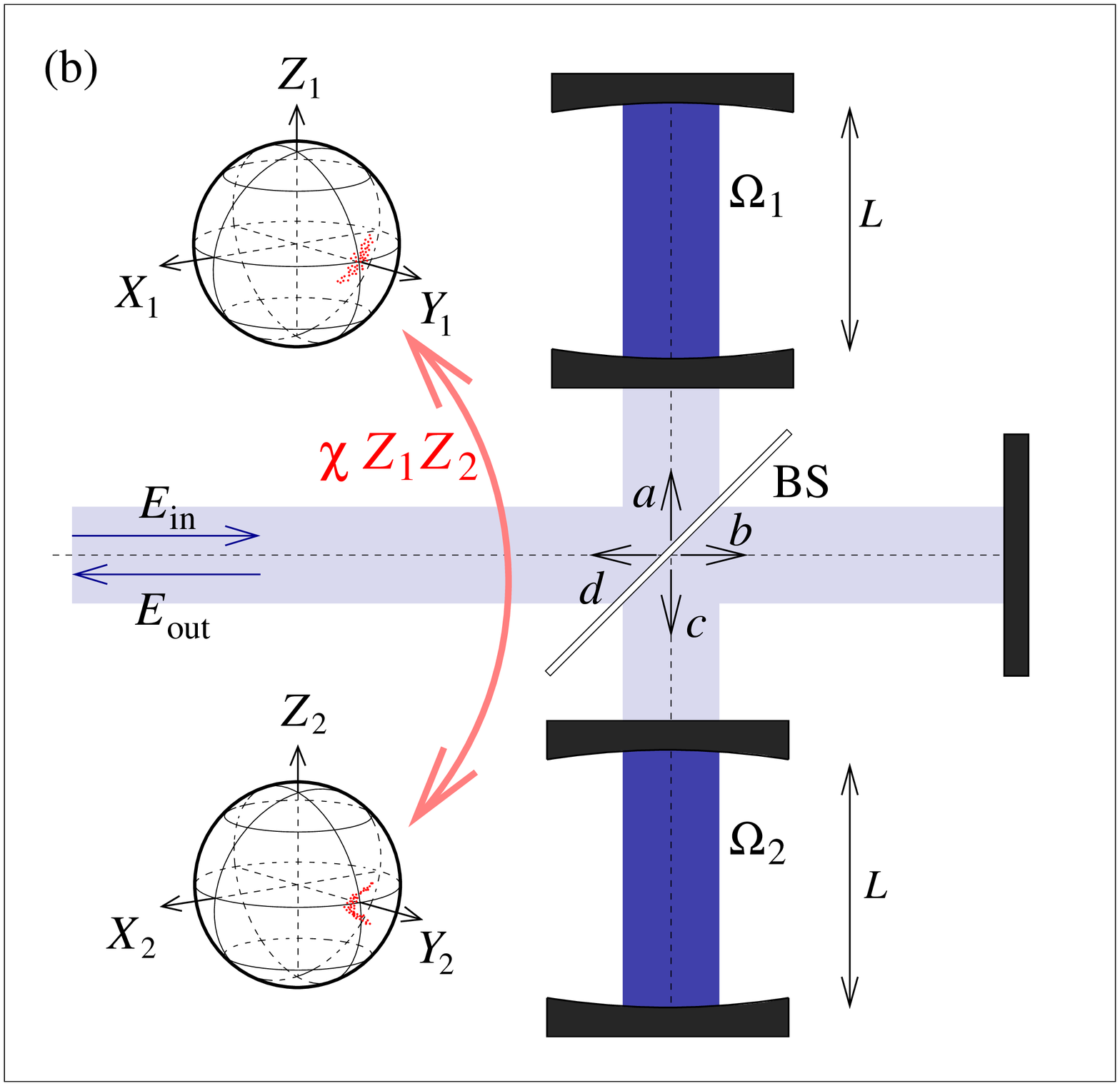,width=0.9\linewidth}}
\caption{\label{f-cavity} (Color online)
(a) Scheme of the resonator with trapped atoms. Red detuned standing wave holds the atoms trapped at locations coinciding with the anti-nodes of the field $E_{\rm cav}$ interacting with the atoms with Rabi frequency $\Omega$. The cavity field frequency is tuned halfway between the transitions $eg_1$ and $eg_2$ such that the phase shift in the cavity is proportional to the difference of atomic numbers in states $g_1$ and $g_2$. The phase in the resonator influences the field intensity inside.
(b) Michelson-like interferometer with two resonators. Difference of atomic numbers in states $g_1$ and $g_2$ corresponds to the spin coordinate $Z$. Phase of each resonator  influences the intensity in both of them. The QND interaction rotates sphere 1 around the $Z_1$ axis in dependence on the value of $Z_2$ and vice versa.
}
\end{figure}
%%%%%%%%%%%%%%%%%  E N D % F I G U R E %%%%%%%%%%%%%%%%%%%%%%

{\em Introduction.---}
Quantum computation with continuous variables (CVs) is an alternative to the computation based on qubits \cite{Lloyd-Braunstein-1999,Braunstein-Loock-2005}. Efficient simulation of quantum processes with dynamical CVs is one of the the main motivations for this approach \cite{Kendon-2010,Georgescu,Deng-2016,Marshall-2015}. A universal CV quantum computer would need (i) a sufficiently large set of single CV modes each of which can be initialized in a suitable quantum state, (ii) a suitable set of single-mode Hamiltonians capable to be combined into more complicated Hamiltonians realizing arbitrary polynomials of the CV, and (iii) a suitable interaction Hamiltonian of different CVs. As shown in \cite{Lloyd-Braunstein-1999}, having in each mode $k$ a conjugate pair of variables $q_k$ and $p_k$ commuting as $[q_k,p_k]=i$, any polynomial Hamiltonian of  $q_k$ and $p_k$ can be constructed if a Hamiltonian of at least third power of   $q_k$ or $p_k$ is available, as well as some simpler Hamiltonians realizing, e.g., displacements or rotations in the phase space. Hamiltonians containing higher powers of $q$ and $p$ are generated by cascaded application of commutators of lower power Hamiltonians. As a possible model of a CV quantum computer, one considers a set of optical modes where Kerr interaction $\propto(q^2+p^2)^2$ plays the role of the higher order Hamiltonian, and beam splitters realize the interaction between different modes. Since the Kerr interaction is typically too weak to be practical for quantum CV operations, alternate schemes have been proposed. These include quantum computing with CV clusters \cite{Menicucci-2006,Gu-2009}, or measurement-based schemes for higher order Hamiltonians such as $\propto q^3$ \cite{Marek-2011,Marshall-2015b}.

Here, a scheme of quasi-CV quantum computation based on collective spins of large ($N \gtrsim 10^3$) atomic samples interacting via optical fields in multi-path interferometers is proposed. Although spin is a discrete variable, for large $N$ and nearly polarized atomic samples the spin components perpendicular to the polarization direction have similar properties as the CVs position $Q$ and momentum $P$ of a harmonic oscillator. Visualizing collective spin states on a Bloch sphere, the computationally relevant states are localized in a confined area where the geometry is close to that of a flat phase space. On the other hand, the curved geometry brings a special advantage in that already quadratic Hamiltonians typically used to generate spin squeezing are sufficient to generate higher power Hamiltonians by commutators. This is achieved by a sequence of rotations (linear Hamiltonian) and squeezing operations (quadratic) which can, in principle, realize Hamiltonians containing, among others, arbitrary powers of the computational variable. Moreover, if the atomic samples are placed in optical resonators mutually coupled to form an interferometer, the off-resonant atom-light interaction can mediate quantum non-demolition (QND) interaction between various samples. By changing resonator lengths and optical phases between the resonators, one can select the modes to interact. Thus, multimode polynomial Hamiltonians can be realized. To realize quantum computation, the system is initialized by squeezing the atomic spins in each resonator, and at the end the results are read-off by measuring the relevant spin components as in the cavity spin squeezing experiments \cite{Leroux-2010a,Hosten-2016}.

%%%%%%%%%%%%%%%%%%%%%%%%%%%%%%%%%%%%%%%%%%%%%%%%%%%%%%
{\em Atoms in a resonator.---}  Based on the idea of atomic spin squeezing by cavity feedback \cite{Schleier-SmithPRA2010,Leroux-2010a}, we first consider a scheme in Fig. \ref{f-cavity}a. 
Incoming laser beam of electric intensity $E_{\rm in}$ is partially reflected from the left cavity mirror and partially enters the cavity. The laser is tuned close to the cavity resonance where the field intensity inside the cavity strongly depends on the optical phase. A large collection of nearly resonant atoms is optically trapped inside the cavity by additional field at anti-nodes of the standing wave $E_{\rm cav}$.  The relevant atomic states are the hyperfine-split states $g_1$ and $g_2$ of the electronic ground state and an electronically excited state $e$. The laser frequency is tuned halfway between the transitions $g_1e$ and $g_2e$ such that the field is detuned by $\Delta$ from  each of them. The presence of an atom in state $g_1$ ($g_2$) changes the optical phase by $\pm\delta \varphi$ respectively, where  $\delta \varphi = \frac{6}{\pi^2} \left( \frac{\lambda}{w}\right)^2 \frac{\Gamma}{\Delta}$ \cite{Supplement}. Here  $\lambda$ is the wavelength,  $w$ is the beam waist, and $\Gamma$ is the optical decay rate from state $e$. 

A collective spin state of the atoms can be expressed in terms of operators $\hat a_{1,2}$  and their Hermitian conjugates, where $\hat a_j^{\dag}$ ($\hat a_j$) creates (annihilates) an atom in state $g_j$, respectively. The total number of atoms is $N=\hat a_1^{\dag}\hat a_1+ \hat a_2^{\dag}\hat a_2$, and the commutation rules are $[\hat a_j, \hat a_k^{\dag}] =\delta_{jk}$. We construct angular momentum-like operators $X$, $Y$, and $Z$ as $X=\frac{1}{2}(\hat a_1^{\dag}\hat a_2 + \hat a_1 \hat a_2^{\dag})$,  $Y=\frac{1}{2i}(\hat a_1^{\dag}\hat a_2 - \hat a_1 \hat a_2^{\dag})$, and $Z=\frac{1}{2}(\hat a_1^{\dag}\hat a_1 -  \hat a_2^{\dag}\hat a_2)$ satisfying the commutation relations $[X,Y]=iZ$,  $[Y,Z]=iX$, and $[Z,X]=iY$. Note that for simplicity we use $X,Y,Z$ rather than the more common notation $J_{x,y,z}$. For states with $|X|,|Z|\ll N$ and $Y \approx N/2$ (near equator, as in Fig. \ref{f-cavity}b), the operators $\tilde q \equiv \sqrt{2/N}Z$ and  $\tilde p \equiv \sqrt{2/N}X$ commute as $[\tilde q,\tilde p] \approx i$ and can be used to simulate the CVs $q$ and $p$.

The optical phase shift in the cavity due to the collective atomic state can be expressed as $\Delta \varphi = 2\delta \varphi Z$.
The cavity phase shift influences the inside field intensity as follows. Assume the left mirror of the cavity has transmissivity $T\ll1$, whereas the right mirror is perfectly reflecting. Assume the loss per one round trip in the cavity is $\epsilon \ll T$. If $\alpha$ describes the optical phase deviation from the center of the resonance line, the field intensity at the anti-nodes in the cavity can be expressed as
$E_{\rm cav}^2 = E_{\rm in}^2 (4/T)[1+\left( \frac{2\alpha}{T} \right)^2]^{-1}$ (see \cite{Supplement} for the derivation).
Expressing the wave number $k=2\pi/\lambda$ as $k=k_0 +\Delta k$ where $k_0L = n\pi$ with $L$ being the cavity length and $n$ integer, the phase deviation is $\alpha=2(L\Delta k + \delta \varphi Z)$. 

The cavity field induces ac Stark shift of the atomic states so that the levels $g_{1,2}$ move apart by $\omega_{\rm ac} = 2\Omega^2/\Delta$  with the Rabi frequency being $\Omega = |E_{\rm cav}| \wp/\hbar$, where $\wp$ is the electric dipole moment of the optical transition. The dipole moment is related to the spontaneous decay rate by
$\Gamma = \frac{1}{4 \pi \epsilon_0} \frac{4 \omega_0^3 \wp^2}{3 \hbar c^3}$ \cite{Scully},
where $\epsilon_0$ is the vacuum permittivity and $\omega_0=k_0c$. The energy of the whole atomic sample is thus changed by $H = 2\hbar \omega_{\rm ac} Z$. Expressing the intensity of the incoming field in terms of the incoming power $P_0=\frac{\pi}{2}\epsilon_0 c w^2 E_{\rm in}^2$ which can be expressed by means of the rate ${\cal R}$ of incoming photons $P_0={\cal R} \hbar \omega_0$, we can write
\begin{eqnarray}
H = \hbar \frac{24}{\pi^2 T}\frac{1}{1+\left(\frac{2\alpha}{T} \right)^2}\left(\frac{\lambda}{w} \right)^2 \frac{\Gamma}{\Delta} Z {\cal R}.
\end{eqnarray}
Linearizing the dependence of $[1+\left( \frac{2\alpha}{T} \right)^2]^{-1}$ on $Z$ for sufficiently large detuning $|\Delta k| \gg \delta \varphi |Z|/L$ one finds
\begin{eqnarray}
H= \hbar \left( \omega Z + \chi Z^2 \right),
\label{eqHam1}
\end{eqnarray}
where
\begin{eqnarray}
\omega &=& \frac{24}{\pi^2T}\frac{1}{1+\left(\frac{4L\Delta k}{T} \right)^2}\left(\frac{\lambda}{w} \right)^2 \frac{\Gamma}{\Delta}  {\cal R}, \\
\chi &=& -\frac{2^7\cdot 3^2}{\pi^4}\frac{1}{T^2}\frac{\frac{4L\Delta k}{T}}{\left[1+\left(\frac{4L\Delta k}{T} \right)^2\right]^2}\left(\frac{\lambda}{w} \right)^4 \left(\frac{\Gamma}{\Delta}\right)^2  {\cal R}.
\end{eqnarray}
Note that a suitable choice of parameters $T$, $L$ and $\Delta k$ is needed so as to ensure both that the interaction is strong enough and that  $\chi$ is independent of $Z$ with a reasonable precision. 
Hamiltonian (\ref{eqHam1}) realizes the one-axis twisting (OAT) scenario of spin squeezing \cite{Kitagawa}.
The sign of  the quadratic term $\chi Z^2$ can be switched by switching the sign of detuning $\Delta k$.  Recently, application of this twist-untwist feature for quantum metrological purposes was proposed \cite{Davis-2016}.

%%%%%%%%%%%%%%%%%  F I G U R E %%%%%%%%%%%%%%%%%%%%%%
\begin{figure}%[h!]
\centerline{\epsfig{file=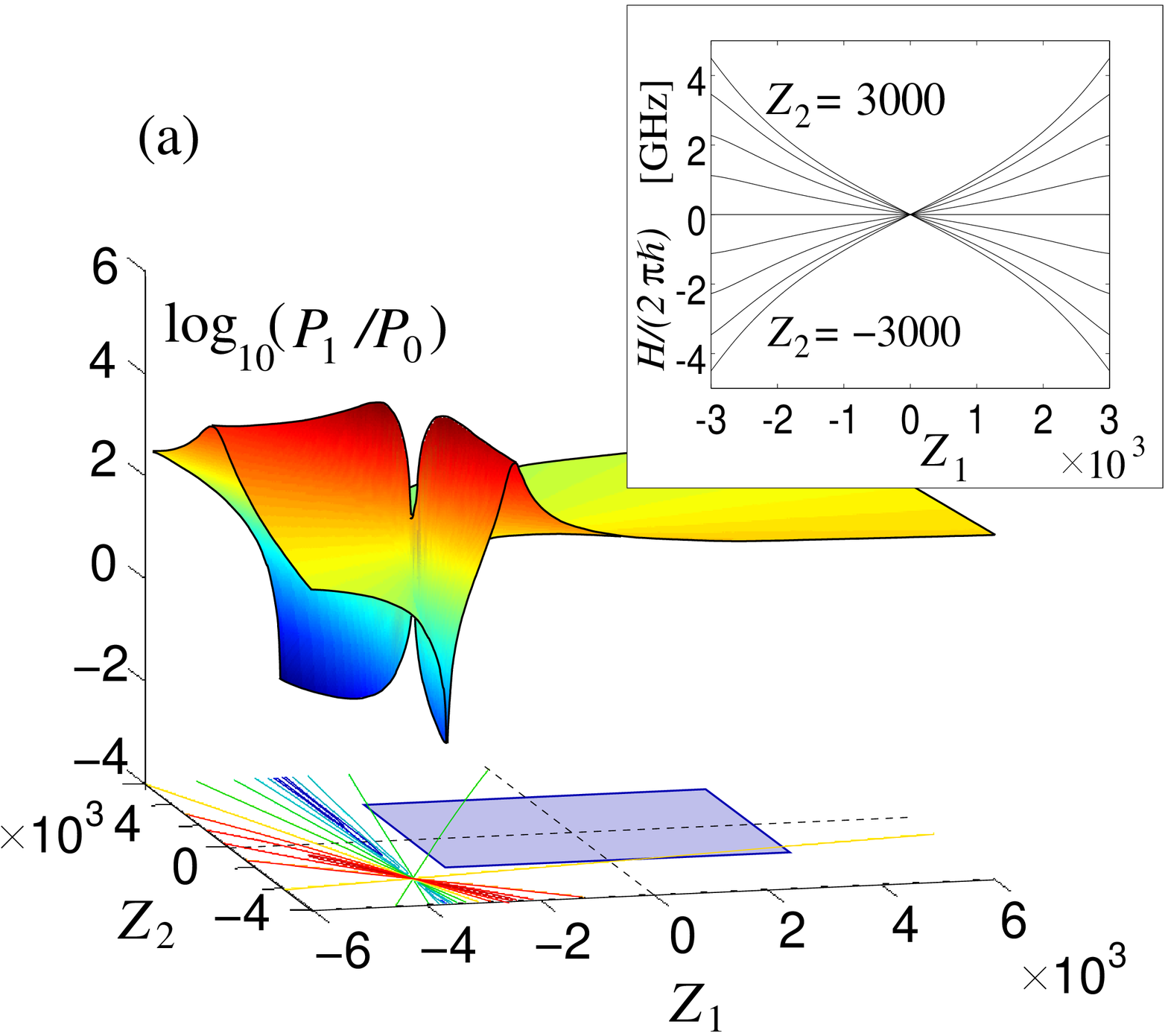,width=1\linewidth}}
\centerline{\epsfig{file=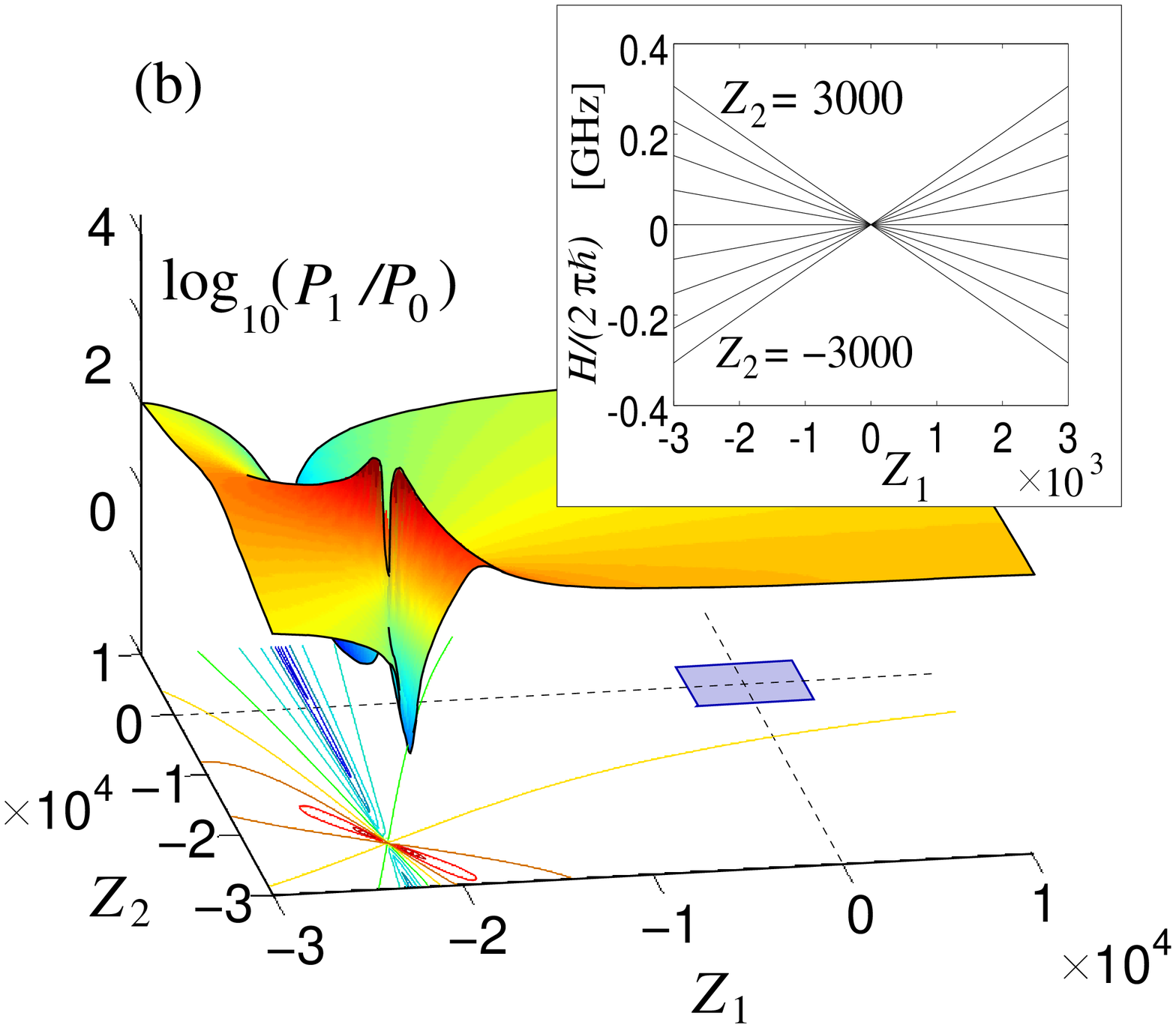,width=1\linewidth}}
\caption{\label{f-Hamilt}
(Color online) Dependence of  power $P_1$ in cavity 1 on the atomic spins $Z_{1,2}$. The lines at the bottom are contours of equal power, the shaded rectangular area of $-3\times 10^3 \leq Z_{1,2} \leq 3\times 10^3$ shows the accessible values with $N=6\times 10^3$ atoms. Insets: resulting interaction Hamiltonian after the four-step sequence described in the text. The lines show dependence of the Hamiltonian on $Z_1$ for 9 equidistant values of $Z_2$ between $\pm 3\times 10^3$. The setup corresponds to that in Fig. \ref{f-cavity}b with $w/\lambda = 100$, cavity mirror transmissivity $T=5\times 10^{-3}$, absorption $\epsilon = 1.2\times 10^{-6}$, cavity length $L=26$ mm, and input power $P_0=12$ nW.
(a) $L\Delta k = 0.08 T$, (b) $L\Delta k = 0.5 T$.
}
\end{figure}
%%%%%%%%%%%%%%%%%  E N D % F I G U R E %%%%%%%%%%%%%%%%%%%%%%

Apart from a quadratic Hamiltonian, one needs also a suitable set of operators linear in the variables $X,Y,Z$. A microwave field off-resonantly coupling states $g_{1,2}$ realizes Hamiltonians proportional to $Z$. A resonant microwave field can realize Hamiltonians proportional to $X\cos \gamma + Y \sin \gamma$ where $\gamma$ is the mutual phase between the microwave and the atomic sample \cite{Leroux-2010a,Schleier-SmithPRL2010}. Alternately, one can also use optical Raman transitions between the spin states.

%%%%%%%%%%%%%%%%%%%%%%%%%%%%%%%%%%%%%%%%%%%%%%%%%%%%%%
{\em Coupling between different atomic samples.---}  First consider a scheme with two cavities in a Michelson-like setup as in Fig. \ref{f-cavity}b. Various modifications are possible, but for concreteness, let us assume cavities in branches $a$ and $c$ with resonant path lengths $2L_ak_0 = 2L_ck_0 = 2\pi n$ and a mirror in path $b$ with $2L_bk_0 = (2n+1)\pi$. In this case a phase shift in one cavity strongly influences the intensity in both cavities. Assuming a sufficiently large detuning 
%changed
$|\Delta k|\gg \epsilon/(2L)$, 
linearization of the phase dependence leads to the Hamiltonian of the form
\begin{eqnarray}
H = \hbar \left[\omega (Z_1 + T_B Z_2) + \chi (Z_1-Z_2)^2 \right], 
\label{eqHamdifquad}
\end{eqnarray} 
where
\begin{eqnarray}
\omega &=&  \frac{2^6\cdot 3}{\pi^2}\frac{R_B}{(1+T_B)^2}\frac{1}{T}\left( \frac{\lambda}{w} \right)^2 \frac{\Gamma}{\Delta}{\cal R} ,\\
\chi &=& - \frac{2^8\cdot 3^2}{\pi^4}\frac{R_B T_B}{(1+T_B)^3}\frac{1}{TL\Delta k}\left( \frac{\lambda}{w} \right)^4 
\left(\frac{\Gamma}{\Delta}\right)^2{\cal R} ,
\end{eqnarray} 
with $T_B=1-R_B$ being the transmissivity of the interferometer beam splitter (see \cite{Supplement} for detailed derivation). Hamiltonian (\ref{eqHamdifquad}) can be straightforwardly used to generate evolution corresponding to the QND Hamiltonian 
\begin{eqnarray}
H_{\rm QND} = -\hbar 2\chi Z_1 Z_2 . 
\label{eqHamQND}
\end{eqnarray} 
This is achieved by a four-step sequence in which rotations of the Bloch spheres  change $Z_j \to -Z_j$ and sign of $\chi$ is changed to the opposite value  in the two steps when exactly one of the coordinates $Z_j$ is changed. This sequence eliminates the linear terms $\propto Z_j$ as well as the quadratic terms  $\propto Z_j^2$ of (\ref{eqHamdifquad}). 

The strength of the interaction $\chi$ can be increased by decreasing the detuning $\Delta k$. This is illustrated in Fig. \ref{f-Hamilt} where the power inside one of the cavities as well as the resulting interaction Hamiltonian are shown for two different values of $\Delta k$. For small $\Delta k$ (Fig. \ref{f-Hamilt}a) the atoms may tune the system close to resonance where the power increases dramatically. This leads to strong interaction, but also to the deformation of the dependence of $H$ on $Z_{1,2}$ beyond the approximation of (\ref{eqHamdifquad}) and (\ref{eqHamQND}). For larger  $\Delta k$ (Fig. \ref{f-Hamilt}b) the Hamiltonian is closer to the bilinear form  (\ref{eqHamQND}), but the interaction is weaker.  The optimal choice of $\Delta k$ will depend on the particular task to be achieved with the interacting atoms.

%%%%%%%%%%%%%%%%%  F I G U R E %%%%%%%%%%%%%%%%%%%%%%
\begin{figure}%[h!]
\centerline{\epsfig{file=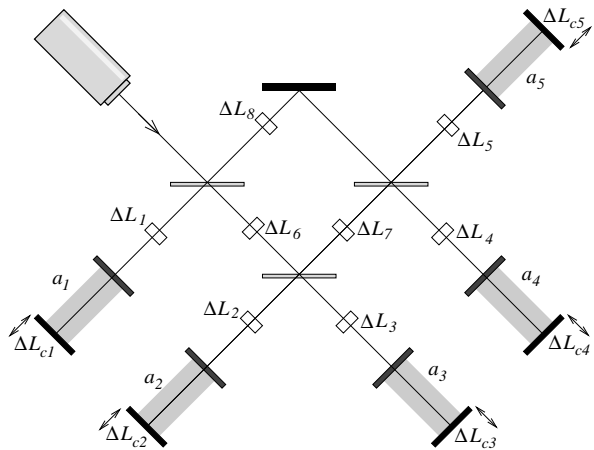,width=0.9\linewidth}}
\caption{\label{f-5cavities}
Scheme of an interferometer with 5 cavities with atomic samples. The interaction between various cavities is switched on and off by changing $\Delta L_{cj}$ to bring the cavities near to resonance or far off-resonance, and by varying the phases in the paths.
}
\end{figure}
%%%%%%%%%%%%%%%%%  E N D % F I G U R E %%%%%%%%%%%%%%%%%%%%%%

The scheme can be scaled up to contain more cavities. The simplest generalization is a  three-cavity scheme where a cavity is placed in each of the $a,b,c$ branches of the interferometer in Fig. \ref{f-cavity}b. By shifting the cavity mirrors, some cavities can be brought sufficiently near to resonance whereas others will be far off-resonant. In the resulting Hamiltonian, only the atomic samples of the nearly resonant cavities will interact. A five-cavity scheme is illustrated in Fig. \ref{f-5cavities}. As checked by numerical simulation of the interferometer, two arbitrary cavities can be brought to interaction whereas the rest of them are switched off.

%%%%%%%%%%%%%%%%%%%%%%%%%%%%%%%%%%%%%%%%%%%%%%%%%%%%%%
{\em Construction of higher power Hamiltonians and functions of the CVs.---} Having the Hamiltonian $\propto Z^2$ with both signs, as well as rotations of the Bloch sphere by linear Hamiltonians, one can construct any quadratic Hamiltonian of $X,Y,Z$. 
In particular, the two-axis countertwisting (TACT) \cite{Kitagawa} Hamiltonians $X^2-Y^2$ or $XY+YX=\frac{1}{2}[(X+Y)^2-(X-Y)^2]$ are built by rotating the Bloch sphere by $\pm \pi/2$ or $\pm \pi/4$ and applying $\pm Z^2$ (for a general treatment of spin squeezing by quadratic Hamiltonians, see \cite{Opatrny-twist}).
Other Hamiltonians can be constructed as commutators of the available operators by the sequence $e^{-iA\Delta t}e^{-iB\Delta t} e^{iA\Delta t}e^{iB\Delta t} = e^{[A,B]\Delta t^2}+ {\cal O}(\Delta t^3)$ and by the Suzuki-Trotter expansion \cite{Trotter,Suzuki76,Hatano-Suzuki}. Using the commutation relations of $X,Y,Z$, one finds, e.g., 
%$X^3 = \frac{i}{4} \left[(X^2-Y^2),(YZ+ZY) \right]+  \frac{i}{4} \left[(XZ+ZX),(XY+YX) \right] + \frac{1}{4}X,$
\begin{eqnarray}
\label{eqx3}
X^3 &=& \frac{i}{4} \left[(Z^2-Y^2),(YZ+ZY) \right] \nonumber \\
 & & +  \frac{i}{4} \left[(XZ+ZX),(XY+YX) \right] + \frac{1}{4}X,
\end{eqnarray}
or a two-mode Hamiltonian
\begin{eqnarray}
\label{eqx3z}
X_1^3 Z_2 &=&\frac{1}{4} X_1 Z_2 + \frac{1}{4}\left[ (Z_1^2 - Y_1^2), \left[ Z_1^2 , X_1 Z_2 \right] \right]
\nonumber \\
&& - \frac{1}{4}\left[X_1 Z_1+Z_1 X_1,\left[X_1^2, Z_1 Z_2 \right] \right],
\end{eqnarray}
that can be useful to construct functions mapping the variables as $(X_1,X_2)\to (X_1, f(X_1)+X_2)$.  By cascading the commutators one can construct Hamiltonians of arbitrary power. More efficient ways of producing various Hamiltonians by using the fact that limited area of the Bloch sphere is used can be found; optimization of the process is in focus of further research.

%%%%%%%%%%%%%%%%%%%%%%%%%%%%%%%%%%%%%%%%%%%%%%%%%%%%%%
{\em Decoherence and losses.---} 
Several challenges have to be addressed to fully utilize the scheme. Losses are inherently connected with the dispersive interaction as $\epsilon \sim N(\lambda/w)^2 (\Gamma/\Delta)^2$ \cite{Supplement}. Decreasing losses thus means also decreasing the strength of the Hamiltonians and thus making the process longer. Therefore, optimization of the interaction strength should be applied to make the process useful. Also, the optical field becomes entangled with the atomic system leading to decoherence: the phase of the outgoing light is influenced by the atomic number inside, and phase of the atomic spins is influenced by the fluctuating light intensity. This problem was studied in detail in \cite{Leroux-2012} and one can anticipate various scenarios to solve it: recycling the light pulses to disentangle them from the atoms, using sub-shot-noise squeezed pulses, or detecting the energy of the outgoing light and considering the atomic state conditioned on the result.

%%%%%%%%%%%%%%%%%%%%%%%%%%%%%%%%%%%%%%%%%%%%%%%%%%%%%%
{\em Discussion and conclusion.---} 
The essential features of the proposed scheme are the possibility to vary the sign of the nonlinearity, to build Hamiltonians of higher powers of the computational CVs out of the quadratic Hamiltonian, and the possibility to couple multiple resonators in interferometric schemes. The seeming contradiction between the possibility to generate higher power Hamiltonians out of quadratic ones and the fact that at least cubic nonlinearity is required in schemes as in \cite{Lloyd-Braunstein-1999} is resolved by considering that the spin operators $X,Y,Z$ themselves are quadratic in the creation and annihilation operators. Thus, the Hamiltonian $Z^2$ contains terms like $\hat a_1^{\dag}\hat a_1 \hat a_2^{\dag}\hat a_2$, i.e., of the cross-Kerr type.

The approach is fully compatible with the scheme of quantum computing with CV clusters \cite{Menicucci-2006,Gu-2009} as all its ingredients are present here: multi-mode squeezed states can be initially prepared by the QND Hamiltonians, and non-Gaussian operations are generated either by $X^3$ and higher order Hamiltonians, or by projective measurements of a suitable non-Gaussian variable. Here, such a measurement can be done by rotating the states close to the pole of the Bloch sphere and then measuring $Z$ which would be analogous to photon counting in optical CV schemes. Note that non-Gaussian features of the detected statistics in atomic spin systems have been used recently for metrology improvement \cite{Strobel-2014}.

The potential of collective spins of atoms in optical resonators for CV quantum computation seems promising taking into account the huge squeezing recently achieved \cite{Hosten-2016}. The scheme is expected to be useful especially for simulation of quantum systems \cite{Kendon-2010,Georgescu,Deng-2016,Marshall-2015}.

%%%%%%%%%%%%%%%%%%%%%%%%%%%%%%%%%%%%%%%%%%%%%%%%%%%%%%
 This work was supported by the Czech Science Foundation, grant No. 17-20479S.

%%%%%%%%%%%%%%%%%%%%%%%%%%%%%%%%%%%%%%%%%%%%%%%%%%%%%%

\begin{widetext}
\newpage

{\bf \Large Supplemental material}
\renewcommand{\theequation}{\thesection.\arabic{equation}}

\vspace{3ex}

This document contains supplemental information for the paper \emph{``Quasi-continuous variable quantum computation with collective spins in multi-path interferometers''} which is referred here to as the ``main text". The document is structured as follows. In Sec. \ref{Phaseshift} the phase shift induced by a single atom in the cavity is derived (result used in section {\em ``Atoms in a resonator''} of the main text). In Sec. \ref{InputOutput} the input-output relations for the cavity and the expression  $E_{\rm cav}^2 = E_{\rm in}^2 (4/T)[1+\left( \frac{2\alpha}{T} \right)^2]^{-1}$ for the field inside the cavity are derived. In Sec. \ref{SpinHamiltonian} the formula for a spin Hamiltonian leading to the results in Eqs. (1)--(4) and (5)--(7) is derived. In Sec. \ref{Absorption} the absorption coefficient $\epsilon$ is estimated, based on off-resonant absorption in the atomic sample. The result is used for estimation of the range of other parameters, in particular the cavity mirror transmittance ($\epsilon \ll T$) and detuning ($\epsilon \ll 2L|\Delta k|$), and in the discussion of the limitations of the scheme. In Sec. \ref{MichelsonInterf} properties of the Michelson-like interferometer with two cavities are derived, with the main results leading to the two-cavity Hamiltonian in Eqs. (5)--(7) of the main text.  In Sec. \ref{Information} information content of the states in qubits is estimated, and in Sec. \ref{Curvature} the influence of the Bloch sphere curvature on the results of quantum simulation is discussed. The results of the last two sections serve for quantitative comparison of the proposed scheme to those of qubit-based computation and of CV computation.

\section{Phase shift by one atom in the cavity}
\label{Phaseshift}
%%%%%%%%%%%%
Assume one atom in an empty cavity resonant for light of frequency $\omega_0 = 2\pi c/\lambda$ with vacuum Rabi frequency $g$ and detuning $\Delta$, 
the frequency shift is $g^2/\Delta$. 
The vacuum Rabi frequency is 
\begin{eqnarray}
g = \frac{{\cal E}_0 \wp}{\hbar},
\end{eqnarray}
where ${\cal E}_0$ is the vacuum electric field and $\wp$ the dipole moment which can be inferred from the relation for the spontaneous decay rate $\Gamma$ (see Ref. [12] of the main text)
\begin{eqnarray}
{\cal E}_0 &=& \sqrt{\frac{\hbar \omega_0}{\epsilon_0 V}},  \\
\Gamma &=& \frac{1}{4 \pi \epsilon_0} \frac{4 \omega_0^3 \wp^2}{3 \hbar c^3}.
\label{EqGammaWp}
\end{eqnarray} 
Here  $V = \pi  \left( \frac{w}{2}\right)^2 L$ is the effective volume of the beam in the cavity, $w$ being the beam waist and $L$  the cavity length. 
From these equations one can find
\begin{eqnarray}
g^2 = \frac{3}{4 \pi}\frac{c\lambda^2 \Gamma}{V} = \frac{3}{8\pi^2} \frac{\lambda^3}{V} \omega_0 \Gamma.
\label{eqg2}
\end{eqnarray}
The change of the wavenumber is then 
\begin{eqnarray}
\delta k = \frac{g^2}{c \Delta},
\end{eqnarray}
and the phase shift per one roundtrip caused by one atom in the cavity of length $L$ is thus 
\begin{eqnarray}
\delta \varphi = 2L\delta k   =  2L \frac{g^2}{c \Delta} = \frac{3}{2\pi} \frac{L\lambda^2}{V} \frac{\Gamma}{\Delta}= \frac{6}{\pi^2} \left( \frac{\lambda}{w}\right)^2 \frac{\Gamma}{\Delta}.
\label{deltavarphi}
\end{eqnarray}

\section{Cavity input-output relations and field inside the cavity}
\label{InputOutput}
%%%%%%%%%%%%%%%%%%%%%%%%%%%%%%%%%%%%%%%%%%%%%%%%%%%%%%

Assume a field $a$ arriving at a cavity mirror with transmissivity $T = t^2$ and reflectivity $R=1-T=r^2$, the geometry being as in Fig 1a of the main text. The field behind the cavity mirror is
\begin{eqnarray}
c = ta_{\rm in} + \sqrt{\eta}r c e^{i(k2L + \varphi)},
\end{eqnarray}
where $L$ is the cavity length, $\eta=1-\epsilon$ with $\epsilon$ being the relative absorption of light in the cavity per one round-trip, and $\varphi$ is an additional phase shift. Introducing $\alpha = 2kL+\varphi ({\rm mod}\ 2\pi)$, for simplicity, we find
\begin{eqnarray}
c &=& \frac{t}{1-\sqrt{\eta}r e^{i\alpha}} a_{\rm in}, \\
a_{\rm out} &=& \frac{\sqrt{\eta}e^{i\alpha} - r}{1-\sqrt{\eta}re^{i\alpha}} a _{\rm in} .
\label{eq-aout-ain}
\end{eqnarray}
For small $\alpha, \epsilon, T \ll 1$ one finds
\begin{eqnarray}
\frac{a_{\rm out}}{a_{\rm in}} \approx \frac{T-\epsilon}{T+\epsilon} + i \frac{4T\alpha}{(T+\epsilon)^2},
\end{eqnarray}
and for $\epsilon \ll T \ll 1$
\begin{eqnarray}
\frac{a_{\rm out}}{a_{\rm in}}   \approx   1-\frac{2\epsilon}{T}  +i \frac{4\alpha}{T}.
\end{eqnarray}
%The field inside the cavity is $P_{\rm cav} = (|c|^2/|a_{\rm in}|^2) P_{\rm in}$  which for small $\alpha$ yields
For the field inside the cavity one can write  $E_{\rm cav}^2 = (|c|^2/|a_{\rm in}|^2) E_{\rm in}^2$   which for small $\alpha$ yields
\begin{eqnarray}
E_{\rm cav}^2 =  \frac{4}{T\left( 1+\frac{\epsilon}{T} \right)^2} \frac{1}{1+\left( \frac{2\alpha}{\epsilon + T} \right)^2}E_{\rm in}^2 \approx   \frac{4}{T} \frac{1}{1+\left( \frac{2\alpha}{ T} \right)^2}E_{\rm in}^2,
\label{inoutcavity}
\end{eqnarray}
where the last approximation is valid for $\epsilon \ll T$.

%%%%%%%%%%%%%%%%%%%%%%%%%%%%%%%%%%%%%%%%%%%%%%%%%%%%%%
\section{Spin Hamiltonian}
\label{SpinHamiltonian}
Consider first a free propagating wave, the transmitted power being 
\begin{eqnarray}
P=\frac{1}{2}\epsilon_0 c E_{\rm max}^2 S,
\end{eqnarray}
where $E_{\rm max}$ is the electric field amplitude and $S$ is the area of the beam. Thus, for the propagating wave one finds
\begin{eqnarray}
E_{\rm max} = \sqrt{\frac{2P}{\epsilon_0 c S}}.
\end{eqnarray}
For a standing wave the amplitude in the anti-nodes is doubled, i.e., 
\begin{eqnarray}
E_{\rm max,standing} = 2\sqrt{\frac{2P}{\epsilon_0 c S}}.
\end{eqnarray}
Atoms located at the standing wave maxima have their frequencies ac-Stark shifted by $\omega_{\rm ac}$, where
\begin{eqnarray}
 \omega_{\rm ac} &=& \frac{\Omega_{\rm max}^2}{\Delta}, \\
 \Omega_{\rm max} &=& \frac{E_{\rm max,standing}\wp}{\hbar} ,
\end{eqnarray}
so that 
\begin{eqnarray}
 \omega_{\rm ac} = \frac{8\wp ^2 P}{\epsilon_0 c \hbar^2 S \Delta}.
\end{eqnarray}
Applying for $\wp$ Eq. (\ref{EqGammaWp}) and for the beam area $S=\pi (w/2)^2$ we get
\begin{eqnarray}
 \omega_{\rm ac} &=& \frac{24}{\pi^2} \left(\frac{\lambda}{w}\right)^2 \frac{\Gamma}{\Delta} \frac{P}{\hbar \omega_0} .
\label{tildeomega}
\end{eqnarray}
If for $N_1$ atoms the detuning is $+\Delta$ and for $N_2$ atoms the detuning is $-\Delta$, the energy change of the system can be described by the Hamiltonian
\begin{eqnarray}
H = 2\hbar \omega_{\rm ac} Z  
&=& \frac{48}{\pi^2} \left( \frac{\lambda}{w} \right)^2 \frac{\Gamma}{\Delta} \frac{P}{\omega_0} Z ,
\end{eqnarray}
where $Z=(N_1-N_2)/2$. If $P$ depends on $Z$, the Hamiltonian becomes nonlinear in $Z$.

%%%%%%%%%%%%%%
\section{Estimation of the absorption coefficient}
\label{Absorption}
Even if one tries to eliminate all losses, there is unavoidable absorption of light interacting with the off-resonant atoms.
The absorption cross section of an atom with the linewidth $\Gamma$ in the field detuned by $\Delta$ from resonance can be expressed as
\begin{eqnarray}
\sigma \approx \frac{6\pi}{k_0^2}\frac{\Gamma^2}{\Gamma^2 + 4 \Delta^2} \approx 
\frac{3}{8\pi}\lambda^2 \left( \frac{\Gamma}{\Delta}\right)^2,
\end{eqnarray}
where the last approximation holds for $\Gamma \ll \Delta$.
If such an atom is sitting on the axis of a beam of waist $w$, the fraction $\sigma/[\pi (w/2)^2]$ of the beam power is absorbed and randomly re-emitted. For $N$ atoms positioned randomly in a cavity, twice this fraction of the circulating light energy is absorbed each round trip per atom, i.e., 
\begin{eqnarray}
\epsilon  &\approx&  2 N \frac{\sigma}{\pi \left(\frac{w}{2}\right)^2} \approx \frac{3}{\pi^2}N \left( \frac{\lambda}{w}\right)^2   \left( \frac{\Gamma}{\Delta}\right)^2 .
\label{eqepsilon}
\end{eqnarray}
For atoms positioned in the anti-nodes, this value is doubled.
For $N=10^4$ of $^{85}$Rb atoms with $\lambda = 780$ nm,  $\Gamma = 2\pi \times 6.06$ MHz, and laser beam of the waist $w=100\ \mu$m detuned by $\Delta = \omega_{HF}/2 = 2\pi \times 3.4$ GHz one finds $\epsilon \approx 1.5\times 10^{-6}$ for the atoms sitting in the anti-nodes.

\section{Michelson-like interferometer with two cavities}
%%%%%%%%%%%%%%%%%%%%%%%%%%%%%%%%%%%%%%%%%%%%%%%%%%%%%%
\label{MichelsonInterf}
\subsection{Input-output relations}
Consider a scheme as in Fig. 1b of the main text. Assume the input amplitude being 1 and the beam splitter transformation matrix 
\begin{eqnarray}
B &=& \left(\begin{array}{cc} 
t_B & ir_B \\ ir_B & t_B
\end{array}
\right)
\end{eqnarray}
with $R_B = r_B^2$, $T_B = t_B^2$, and $R_B+T_B=1$.
The amplitudes in the four outputs of the beam splitter can be calculated from
\begin{eqnarray}
\label{eqabcd1}
a &=& ir_B + t_B c e^{i 2L_ck} f_c, \\
b &=& t_B + ir_B c e^{i 2L_ck} f_c, \\
c &=& t_B a e^{i 2L_ak} f_a  + ir_B b e^{i 2L_bk}, \\
d &=& t_B b e^{i 2L_bk}  + ir_B a e^{i 2L_ak} f_a,
\label{eqabcd2}
\end{eqnarray}
where
\begin{eqnarray}
f_a &=& \frac{\sqrt{1-\epsilon}e^{i\alpha_a}-\sqrt{1-T}}{1-\sqrt{1-\epsilon}\sqrt{1-T}e^{i\alpha_a}}
\end{eqnarray}
is the input-output relation of the cavity in path $a$ (cf. Eq. (\ref{eq-aout-ain}))
and similarly for $f_c$. Each cavity has length $L$, and the distances between the beam splitter and the mirror in paths $a,b,c$ are $L_{a,b,c}$.
Solving the set (\ref{eqabcd1})--(\ref{eqabcd2}) one gets
\begin{eqnarray}
a &=& ir_B \frac{1+e^{i2(L_c+L_b)k}f_c}{1+r_B^2 e^{i2(L_c+L_b)k}f_c - t_B^2  e^{i2(L_a+L_c)k}f_a f_c }, \nonumber \\
b &=& t_B \frac{1-e^{i2(L_a+L_c)k}f_a f_c}{1+r_B^2 e^{i2(L_c+L_b)k}f_c - t_B^2  e^{i2(L_a+L_c)k}f_a f_c }, \nonumber \\
c &=& ir_B t_B \frac{e^{i2L_bk}+ e^{i2L_ak}f_a}{1+r_B^2 e^{i2(L_c+L_b)k}f_c - t_B^2  e^{i2(L_a+L_c)k}f_a f_c }, \nonumber \\
d &=& \frac{t_B^2 e^{i2L_bk}- r_B^2 e^{i2L_ak} f_a -e^{i2(L_a+L_b+L_c)k}f_a f_c}{1+r_B^2 e^{i2(L_c+L_b)k}f_c - t_B^2  e^{i2(L_a+L_c)k}f_a f_c }. \nonumber \\
\label{eqabcd4}
\end{eqnarray}
From these equations one gets the intensities
\begin{eqnarray}
\label{eqa2rb}
|a|^2 &=& R_B \frac{1+|f_c|^2 + 2|f_c|\cos \delta_c}{J}, \\
|b|^2 &=& T_B \frac{1+|f_a|^2|f_c|^2 - 2|f_a||f_c|\cos ( \delta_a+ \delta_c)}{J}, \\
|c|^2 &=& R_B T_B \frac{1+|f_a|^2 + 2|f_a|\cos \delta_a}{J}, 
\label{eqc2rb}\\
|d|^2 &=& \left[T_B^2 +R_B^2|f_a|^2+|f_a|^2|f_c|^2  - 2T_B|f_a| |f_c|\cos (\delta_a+\delta_c) 
 \right. \nonumber \\
&& \left.
- 2R_BT_B|f_a|\cos \delta_a + 2R_B|f_a|^2 |f_c| \cos \delta_c  \right]/J,
\label{eqd2rb}
\end{eqnarray}
where
\begin{eqnarray}
J &=& 1+R_B^2 |f_c|^2 + T_B^2 |f_a|^2|f_c|^2 + 2 R_B |f_c| \cos \delta_c 
%\nonumber \\
%&& 
- 2T_B |f_a| |f_c|  \cos (\delta_a+\delta_c) -  2 R_B T_B |f_a| |f_c|^2 \cos \delta_a , 
%\nonumber \\
\label{denomJ}
\end{eqnarray}
and 
\begin{eqnarray}
\delta_a &\equiv& \gamma_a + 2(L_a-L_b)k, \\
\delta_c &\equiv& \gamma_c + 2(L_c+L_b)k, \\ 
\end{eqnarray}
with
\begin{eqnarray}
f_{a,c} &\equiv& |f_{a,c}| e^{i\gamma_{a,c}} .
\end{eqnarray}
As can be checked, for $|f_a| = |f_c|=1$ (no losses, $\epsilon =0$) one gets $|d|^2=1$, i.e., energy conservation.

%%%%%%%%%%%%%%%%
\subsection{Estimation of losses in resonance}

Resonance effects occur when the denominator $J$ is small. In the lossless case  $|f_a| = |f_c|=1$ one finds that $J=0$ for $\delta_a=\pm \pi$ and  $\delta_c=\pm \pi$. For small losses with $\epsilon \ll T$ and $T, \gamma_{a,c} \ll 1$ one can expand the expressions as powers of $\epsilon/T$ and $\gamma_{a,c}$,  
\begin{eqnarray}
\label{apf1}
|f_{a,c}| &\approx& 1-2\frac{\epsilon}{T}+\frac{\epsilon}{2T}\gamma_{a,c}^2+2\left(\frac{\epsilon}{T} \right)^2 -\left(\frac{\epsilon}{T} \right)^3, \\
|f_{a,c}|^2 &\approx& 1-4\frac{\epsilon}{T}+\frac{\epsilon}{T}\gamma_{a,c}^2+8\left(\frac{\epsilon}{T} \right)^2 - 10\left(\frac{\epsilon}{T} \right)^3, \\
|f_{a}||f_{c}| &\approx& 1-4\frac{\epsilon}{T}+\frac{\epsilon}{2T}\left( \gamma_{a}^2 +\gamma_{c}^2\right) +8\left(\frac{\epsilon}{T} \right)^2 - 10\left(\frac{\epsilon}{T} \right)^3, \\
|f_{a}|^2|f_{c}| &\approx& 1-6\frac{\epsilon}{T}+\frac{\epsilon}{2T}\left(2 \gamma_{a}^2 +\gamma_{c}^2\right) +18\left(\frac{\epsilon}{T} \right)^2 - 35\left(\frac{\epsilon}{T} \right)^3, \\
|f_{a}||f_{c}|^2 &\approx& 1-6\frac{\epsilon}{T}+\frac{\epsilon}{2T}\left(\gamma_{a}^2 +2\gamma_{c}^2\right) +18\left(\frac{\epsilon}{T} \right)^2 - 35\left(\frac{\epsilon}{T} \right)^3, \\
|f_{a}|^2|f_{c}|^2 &\approx& 1-8\frac{\epsilon}{T}+\frac{\epsilon}{T}\left(\gamma_{a}^2 +\gamma_{c}^2\right) +32\left(\frac{\epsilon}{T} \right)^2 - 84\left(\frac{\epsilon}{T} \right)^3.
\label{apf2}
\end{eqnarray}
Expressing $|d|^2 = C/J$ with $C$ and $J$ expanded up to the third power of $\epsilon/T$, one finds
\begin{eqnarray}
C &\approx& 4(1+T_B)^2  \left(\frac{\epsilon}{T} \right)^2 - 8(3+4T_B+T_B^2) \left(\frac{\epsilon}{T} \right)^3 , \\
J &\approx& 4(1+T_B)^2  \left(\frac{\epsilon}{T} \right)^2 - 8(1+4T_B+3T_B^2) \left(\frac{\epsilon}{T} \right)^3 .
\end{eqnarray}
This leads to
\begin{eqnarray}
|d|^2 &\approx& 1-4\frac{1-T_B}{1+T_B}\frac{\epsilon}{T},
\label{d2apr}
\end{eqnarray}
which can be used to estimate the total losses in the interferometer.

%%%%%%%%%%%%%%%%
\subsection{Estimation of the inter-cavity intensity dependence on phases}

The intensities inside the cavities can be expressed using Eq. (\ref{inoutcavity}) where we assume that the phases $\alpha_{1,2}$ are changed in an interval much narrower than $T$ so that $P_{\rm cav}\approx \frac{4}{T}P_{\rm in}$ and for $P_{\rm in}$ we use Eqs. (\ref{eqa2rb}) and (\ref{eqc2rb}) multiplied with $P_0$, i.e., the field incoming to the system. We thus get
\begin{eqnarray}
\label{P1pom}
P_1 &\approx& \frac{4}{T} R_B\frac{1+|f_c|^2 +2|f_c|\cos \delta_c}{J}P_0, \\
P_2 &\approx& \frac{4}{T} R_B T_B \frac{1+|f_a|^2 +2|f_a|\cos \delta_a}{J}P_0.
\label{P2pom}
\end{eqnarray}
We use
\begin{eqnarray}
\delta_a &=& \gamma_a + 2(L_a-L_b)k \ ({\rm mod}\ 2\pi) \equiv \frac{4\alpha_1}{T} + 2(L_a-L_b)(k_0+\Delta k) \ ({\rm mod}\ 2\pi) , \\
\delta_c &=& \gamma_c + 2(L_c+L_b)k \ ({\rm mod}\ 2\pi)  \equiv \frac{4\alpha_2}{T} + 2(L_c+L_b)(k_0+\Delta k) \ ({\rm mod}\ 2\pi) ,
\end{eqnarray}
and 
\begin{eqnarray}
\alpha_{1,2} &=& \varphi_{1,2}+2L\Delta k, \\
L_a k_0 &=& 2\pi n_a, \\
L_c k_0 &=& 2\pi n_c, \\
L_b k_0 &=& 2\pi n_b + \frac{\pi}{2}, \\
\end{eqnarray}
$n_{a,b,c}$ being integers, so that we arrive at
\begin{eqnarray}
\delta_a &=& \frac{4}{T}\left( \varphi_1 + 2L \Delta k\right) - \pi - \pi\frac{\Delta k}{k_0} \approx \frac{4}{T}\left( \varphi_1 + 2L \Delta k\right) - \pi, \\
\delta_c &=& \frac{4}{T}\left( \varphi_2 + 2L \Delta k\right) + \pi + 8n\pi\frac{\Delta k}{k_0} \approx \frac{4}{T}\left( \varphi_2 + 2L \Delta k\right) + \pi,
\end{eqnarray}
where we have neglected the last terms as we assume $n_{a,b,c}\pi/k_0 \ll L/T$, i.e., $T\ll L/L_{a,b,c}$. In this case we find
\begin{eqnarray}
1+|f_a|^2 +2|f_a|\cos \delta_a &\approx& 4\left(\frac{\epsilon}{T} \right)^2+\left(\frac{4}{T} \right)^2 \left(\varphi_1 +2L\Delta k \right)^2, \\
1+|f_c|^2 +2|f_c|\cos \delta_c &\approx& 4\left(\frac{\epsilon}{T} \right)^2+\left(\frac{4}{T} \right)^2 \left(\varphi_2 +2L\Delta k \right)^2,
\end{eqnarray}
and
\begin{eqnarray}
J &\approx & 4 (1+T_B)^2 \left(\frac{\epsilon}{T} \right)^2 + \left(\frac{4}{T} \right)^2 \left[R_B(\varphi_2 +2L\Delta k)^2 +T_B (\varphi_1+ \varphi_2 +4L\Delta k)^2
- R_B T_B (\varphi_1 +2L\Delta k)^2 \right] \\
&=& 4 (1+T_B)^2 \left(\frac{\epsilon}{T} \right)^2 + \left(\frac{4}{T} \right)^2 \left[\varphi_2 + T_B \varphi_1 +(1+T_B)2L\Delta k \right]^2.
\end{eqnarray}
Combined with (\ref{P1pom}) and (\ref{P2pom}), these equations yield
\begin{eqnarray}
P_1 &\approx& \frac{4}{T}\frac{R_B}{(1+T_B)^2}
\frac{\epsilon^2+4\left(\varphi_2 +2L\Delta k \right)^2}{\epsilon^2+\frac{4}{(1+T_B)^2} \left[\varphi_2 + T_B \varphi_1 +(1+T_B)2L\Delta k \right]^2}P_0, \\
P_2 &\approx& \frac{4}{T}\frac{R_B T_B}{(1+T_B)^2}
\frac{\epsilon^2+4\left(\varphi_1 +2L\Delta k \right)^2}{\epsilon^2+\frac{4}{(1+T_B)^2} \left[\varphi_2 + T_B \varphi_1 +(1+T_B)2L\Delta k \right]^2}P_0 .
\end{eqnarray}
For sufficiently large detuning  $|\Delta k| \gg \epsilon/(2L)$ one can write
\begin{eqnarray}
P_1 &\approx& \frac{4R_B}{T}
\frac{(\varphi_2+2L\Delta k)^2}{\left[\varphi_2+T_B\varphi_1 +(1+T_B)2L\Delta k\right]^2}P_0, \\
P_2 &\approx& \frac{4R_BT_B}{T}
\frac{(\varphi_1+2L\Delta k)^2}{\left[\varphi_2+T_B\varphi_1 +(1+T_B)2L\Delta k\right]^2}P_0, 
\end{eqnarray}
which for $\varphi_{1,2} \ll L\Delta k$ can be linearized as
\begin{eqnarray}
P_1 &\approx& \frac{4}{T}  \frac{R_B}{(1+T_B)^2}
\left[1 + \frac{T_B (\varphi_2-\varphi_1)}{(1+T_B)L\Delta k} \right] P_0, \\
P_2 &\approx&  \frac{4}{T}  \frac{R_B T_B}{(1+T_B)^2}
\left[1 + \frac{\varphi_1-\varphi_2}{(1+T_B)L\Delta k} \right] P_0 .
\end{eqnarray}
Expressing $\varphi_{1,2} = 2\delta \varphi Z_{1,2}$ with $\delta \varphi$ of (\ref{deltavarphi}) and writing the total Hamiltonian as 
\begin{eqnarray}
H = 2\hbar \left(\omega_{{\rm ac} 1} Z_1 +   \omega_{{\rm ac} 2}  Z_2 \right)
\end{eqnarray}
with $\omega_{{\rm ac} 1,2}$ given by (\ref{tildeomega}), we arrive at Eqs. (5)--(7) of the main text.

\section{Information in the quasi-continuous variables}
\label{Information}
%%%%%%%%%%%%%%%%%%%%%%%%%%%%%%%%%%%%%%%%%%%%%%%%%%%%%%
Symmetric states of $N$ two-level atoms form a $(N+1)$ dimensional Hilbert space that can, in principle, encode $\log_2 (N+1)$ qubits. If one wants to use the system for continuous-variable simulation, only a part of this number is available. To estimate the reduction,
assume that the states of interest are located on the Bloch sphere in an area of radius $r \frac{N}{2}$ much smaller than the  Bloch sphere radius $N/2$, i.e., $r \ll 1$.
Assume that the information is encoded into spin-squeezed states of principal half-widths (square roots of variances) $\Delta n_{\pm}$ where the larger half-width is $\Delta n_+ = r \frac{N}{2}$, and the product of the half-widths is the variance of the spin coherent state, $\Delta n_+ \Delta n_- = \frac{N}{4}$. Thus, the smaller half-width is $\Delta n_- = \frac{1}{2r}$, and the number of states that can be encoded is $\approx \frac{\Delta n_+}{\Delta n_-} = r^2 N$. Thus, the information corresponds to $\log_2 (r^2 N)$, i.e., the reduction is $\approx 2\log_2 r$ bits (more details on the amount of information carried by continuous variables can be found, e.g., in \cite{Braunstein-Loock-2005}). Using these relations, one finds that
\begin{eqnarray}
N_{\rm bits} = -\frac{\rm Sq}{10\log_{10} 2},
\end{eqnarray}
where Sq refers to the squeezing in decibels (negative Sq means squeezing below the standard quantum limit) and $N_{\rm bits}$ is the number of qubits encoded in these states.

As an illustration, assume $N=6000$ atoms and $r = 0.05$. This corresponds to $\log_2 6000 \approx$ 12.5-qubit equivalent of the total number of symmetric states reduced to 3.9 qubits encoded in the 15 states located in the $r N/2=150$ radius area of the Bloch sphere. These states are squeezed by $10 \log_{10}[\Delta n_-^2/(N/4)]$ dB $\approx -11.7$ dB. 

As another example, consider the recent record in spin squeezing of $-$20 dB in a sample of $N=5\times 10^{5}$ atoms \cite{Hosten-2016}. Using these states for continuous variable simulation, one finds $r = 0.014$ which would allow to work with $\approx 100$ different states, corresponding to $\log_2 100 \approx 6.6$ qubits in one sample. Note that using all the symmetrical states of the $5\times 10^{5}$-sample would correspond to approximately $18.9$ qubits.

\section{Influence of the Bloch sphere curvature on the squeezed state overlap}
\label{Curvature}
%%%%%%%%%%%%%%%%%%%%%%%%%%%%%%%%%%%%%%%%%%%%%%%%%%%%%%
For large $N$ and weakly excited states (i.e., states not too distant from a chosen spin coherent state) the system dynamics are virtually the same as the dynamics of a harmonic oscillator having a flat phase space. 
%This is the case, e.g., of experiments with the atomic quantum memory for squeezed light \cite{Appel-2008}. 
On the other hand, strongly squeezed atomic spin states differ from their harmonic oscillator counterparts because of the finite-dimensional Hilbert space and curved phase space (Bloch sphere). 
%The corresponding features of the resulting states have been utilized, e.g., in \cite{Strobel-2014}.
For simulating continuous variable systems, the resulting states should be close to those of the flat phase space, although advantage of the curved Bloch sphere is taken during the computation process. 

There are various possibilities to quantify the influence of the curvature of the Bloch sphere on the states of interest. As an example, one can compare overlaps of mutually shifted squeezed coherent states. First consider two Gaussian states $|\psi_{1,2}\rangle$ of a harmonic oscillator. Let the states be located parallel to each other: the mean values of their quadratures are $\langle q_1 \rangle$, $\langle q_2\rangle$, and $\langle p_1\rangle = \langle p_2\rangle = 0$, and their standard deviations are $\Delta q_1 = \Delta q_2 \equiv \Delta q$ and $\Delta p_1 = \Delta p_2 \equiv \Delta p$ with $\Delta q\Delta p = \frac{1}{2}$ as for the minimum uncertainty states. If the mutual distance of the Gaussians is a fixed multiple of their widths, $\langle q_2 \rangle -\langle q_1 \rangle = \xi \Delta q$, the overlap of the states $|\langle \psi_1|\psi_2 \rangle|^2$ only depends on $\xi$ as  $|\langle \psi_1|\psi_2 \rangle|^2 = \exp (-\xi^2/4)$. 

The situation is changed in a curved phase space. Consider two spin squeezed states centered on the equator and stretched along the meridians. Let the distance of their centers be a fixed multiple of their width, and let us vary their extension along the meridian. For small extensions the situation is similar as in the flat phase space, but when the stretching becomes comparable to the size of the Bloch sphere, one can observe the effect of meridians approaching each other near the poles. The results can be seen in Fig. \ref{f-overlaps}. One can observe decreasing or increasing overlap of the states due to the interference of the approaching parts in the phase space. As can be seen, the results are very close to those in the flat phase space provided that the relative extension of the state to the Bloch sphere radius $r$ is $r \lesssim 0.1$.

%%%%%%%%%%%%%%%%%  F I G U R E %%%%%%%%%%%%%%%%%%%%%%
\begin{figure}%[h!]
\centerline{\epsfig{file=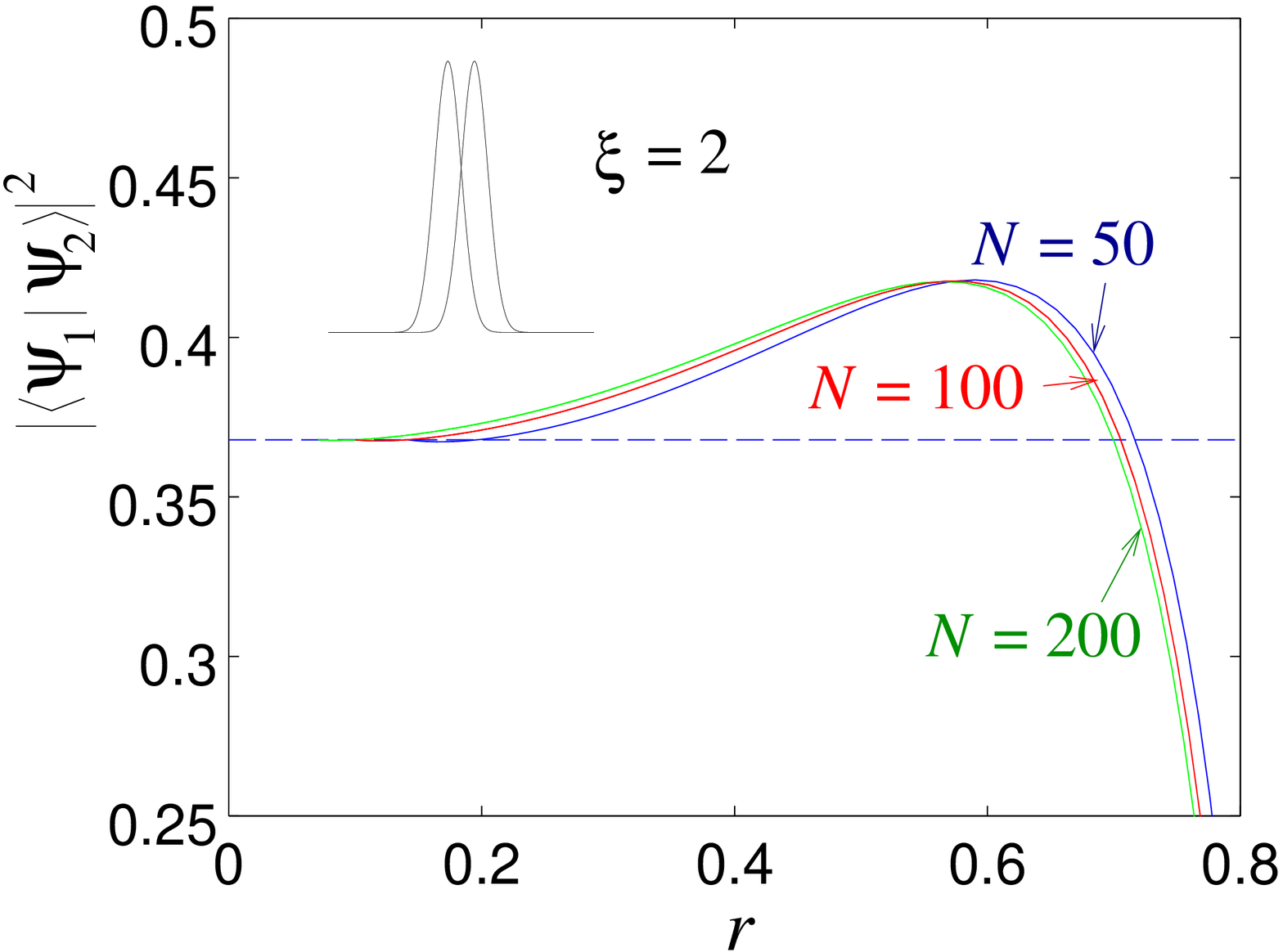,width=0.3\linewidth}
\epsfig{file=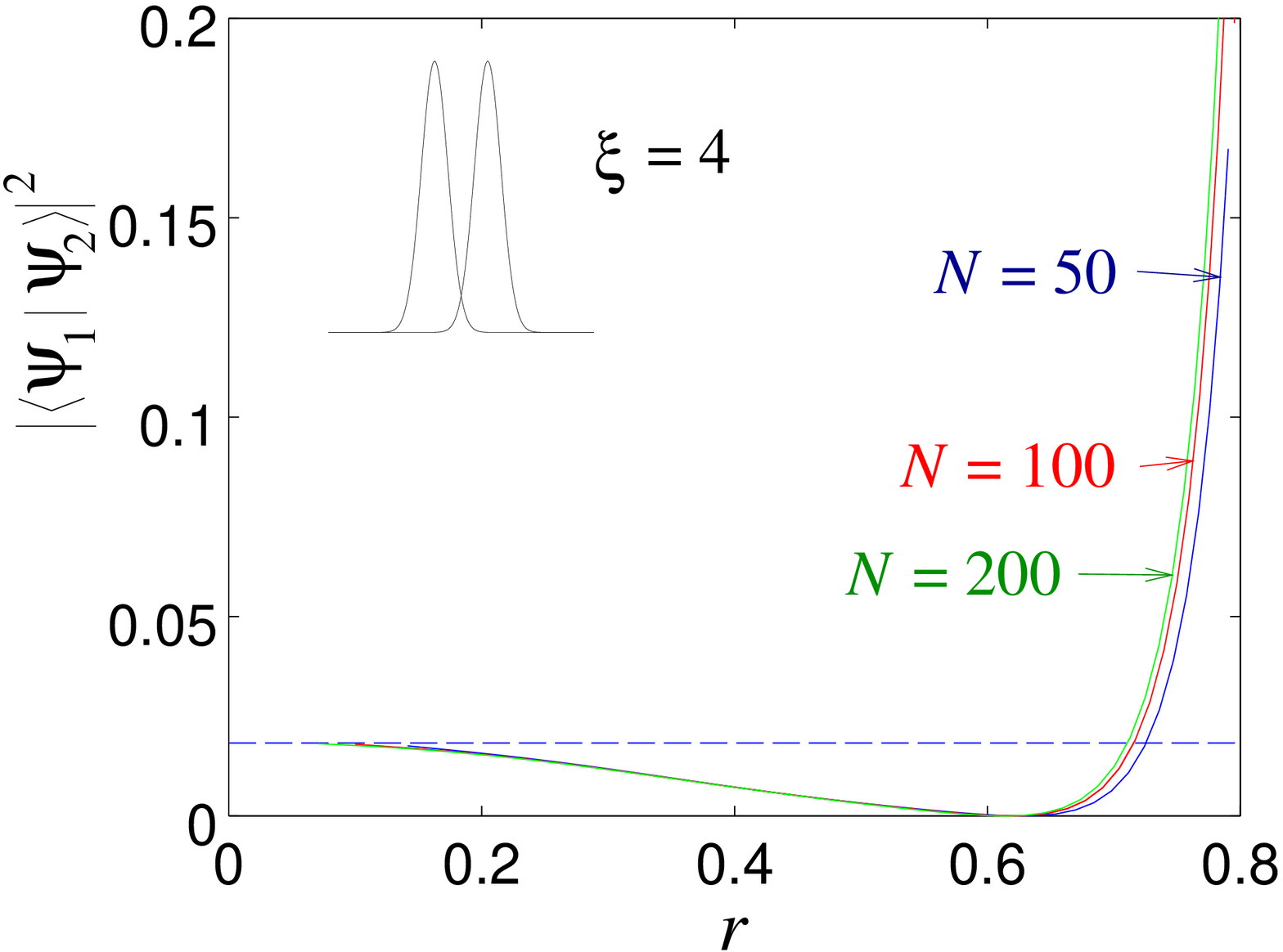,width=0.3\linewidth}
\epsfig{file=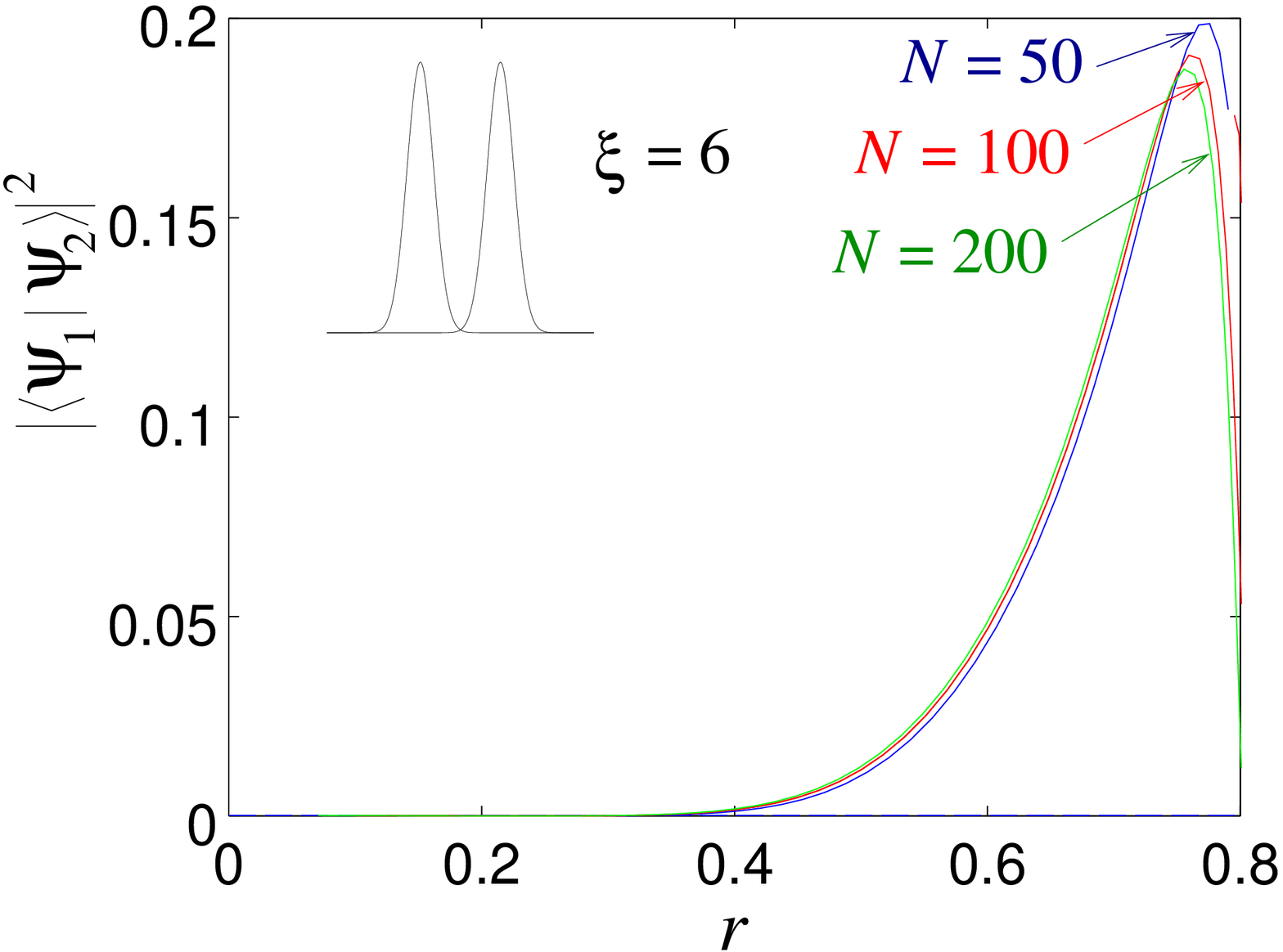,width=0.3\linewidth}}
\caption{\label{f-overlaps}
Dependence of the overlap of two spin-squeezed states on their relative size with respect to the Bloch sphere radius calculated for various atomic numbers $N$. The states are prepared by squeezing the spin coherent state $|\psi_0\rangle = \exp (-i Y \pi/2)|0\rangle$ localized at the equator and polarized along the $X$ axis, where $|0\rangle$ is the vacuum state (all atoms in state $g_2$). Squeezing is achieved by acting with the TACT Hamiltonian $YZ+ZY$ for time $t$, i.e., the resulting state is $|\psi_1\rangle = \exp [-i(YZ+ZY)t]|\psi_0\rangle$. This state is polarized in the $X$ direction and is squeezed in $Y$ and stretched in $Z$ with variances $\langle \psi_1|Y^2|  \psi_1\rangle = \Delta n_-^2$ and $\langle \psi_1|Z^2|  \psi_1\rangle = \Delta n_+^2$. Parameter $r$ quantifies the relative extension of the state $\Delta n_+$ with respect to the Bloch sphere radius $N/2$ as $r = 2\Delta n_+/N$. State $|\psi_2\rangle$ is produced by rotating  $|\psi_1\rangle$ around $Z$ by angle $\phi$ such that the state is displaced by a $\xi$-multiple of its squeezed half-width $\Delta n_-$, i.e., $|\psi_2\rangle = \exp (i Z \phi)|  \psi_1\rangle$, where $\phi = 2\xi \Delta n_-/N$. The insets show the mutual position of two pure Gaussian states of a harmonic oscillator displaced by the same multiple $\xi$ of their quadrature half-widths. The dashed line shows the overlap of these harmonic oscillator states, $|\langle \psi_1|\psi_2 \rangle|^2 = \exp (-\xi^2/4)$. One can see that for states localized in a relatively small fraction of the Bloch sphere $r \lesssim 0.1$ the results of the collective spins are very close to those of harmonic oscillators.
}
\end{figure}
%%%%%%%%%%%%%%%%%  E N D % F I G U R E %%%%%%%%%%%%%%%%%%%%%%

\end{widetext}


\begin{thebibliography}{99}


\bibitem{Lloyd-Braunstein-1999}
S. Lloyd and S. L. Braunstein,
\textit{Quantum Computation over Continuous Variables.}
Phys. Rev. Lett.  {\bf 82,}  1784 (1999).



\bibitem{Braunstein-Loock-2005}
S. L. Braunstein and P. van Loock, 
\textit{Quantum information with continuous variables.}
Rev. Mod. Phys.  {\bf 77,}  5013 (2005).
% contains Quantum Computation with Continuous Variables A. Universal quantum computation B. Extension of the Gottesman-Knill theorem
% Lloyd-Braunstein model in a bit more detail

\bibitem{Kendon-2010}
V. M. Kendon, K. Nemoto, and W. J. Munro,
\textit{Quantum analogue computing.}
Phil. Trans. R. Soc. A {\bf 368,} 3609 (2010). 

\bibitem{Georgescu}
I. M. Georgescu, S. Ashhab, and F. Nori,
\textit{Quantum simulation.}
Rev. Mod. Phys. {\bf 86,} 153 (2014).



\bibitem{Marshall-2015}
K. Marshall, R. Pooser, G. Siopsis, and C.  Weedbrook,
\textit{Quantum simulation of quantum field theory using continuous variables.}
Phys. Rev. A {\bf 92,} 063825 (2015).

\bibitem{Deng-2016}
X. Deng, S. Hao, H. Guo, C. Xie, and X. Su,
\textit{Continuous variable quantum optical simulation for time evolution of quantum harmonic oscillators.}
Sci. Rep.  {\bf 6,} 22914 (2016).

\bibitem{Menicucci-2006}
N. C. Menicucci, P. van Loock, M. Gu, C. Weedbrook, T. C. Ralph, and M. A. Nielsen,
\textit{Universal quantum computation with continuous-variable cluster states.}
Phys. Rev. Lett.  {\bf  97,} 110501 (2006).


\bibitem{Gu-2009}
M. Gu,  C. Weedbrook, N. C. Menicucci,  T. C. Ralph,  and P. van Loock,
\textit{Quantum computing with continuous-variable clusters.}
Phys. Rev. A {\bf 79,}  062318 (2009).


\bibitem{Marek-2011}
P. Marek, R. Filip, and A. Furusawa,
\textit{Deterministic implementation of weak quantum cubic nonlinearity.}
Phys. Rev. A {\bf 84,}  053802 (2011).


\bibitem{Marshall-2015b}
K. Marshall, R. Pooser, G. Siopsis, and C.  Weedbrook,
\textit{Repeat-until-success cubic phase gate for universal continuous-variable quantum computation.}
Phys. Rev. A {\bf 91,}  032321 (2015).

\bibitem{Leroux-2010a}
I. D. Leroux, M. H. Schleier-Smith,  and V. Vuleti\'{c},
\textit{Implementation of cavity squeezing of a collective atomic spin.}
Phys. Rev. Lett.  {\bf 104,}   073602 (2010).

\bibitem{Hosten-2016}
O. Hosten, N. J. Engelsen, R. Krishnakumar, and M. A. Kasevich,
\textit{Measurement noise 100 times lower than the quantum-projection limit using entangled atoms.}
Nature Phys. {\bf 529,}  505 (2016).


\bibitem{Schleier-SmithPRA2010}
M. H. Schleier-Smith, I. D. Leroux, and V. Vuleti\'{c},
\textit{Squeezing the collective spin of a dilute atomic ensemble by cavity feedback.}
Phys. Rev. A  {\bf 81,} 021804(R) (2010).


\bibitem{Supplement}
See the supplemental material.

\bibitem{Scully}
M. O. Scully and M. S. Zubairy,
\textit{Quantum optics.} Cambridge University Press 1997.

\bibitem{Kitagawa}
M. Kitagawa and M. Ueda,
\textit{Squeezed spin states.}
Phys. Rev. A {\bf 47,} 5138 (1993).

\bibitem{Davis-2016}
E. Davis, G. Bentsen, and   M. H. Schleier-Smith,  
\textit{Approaching the Heisenberg limit without single-particle detection.}
Phys. Rev. Lett.  {\bf 116,}  053601(2016).



\bibitem{Schleier-SmithPRL2010}
M. H. Schleier-Smith, I. D. Leroux, and V. Vuleti\'{c},
\textit{States of an ensemble of two-level atoms with reduced quantum uncertainty.}
Phys. Rev. Lett.  {\bf 104,}  073604 (2010).


\bibitem{Opatrny-twist}
T. Opatrn\'{y},
\textit{Twisting tensor and spin squeezing.}
Phys. Rev. A {\bf 91}, 053826 (2015)

\bibitem{Trotter}
H. F. Trotter,
\textit{On the product of semi-groups of operators.} 
Proceedings of the American Mathematical Society {\bf 10,}
 pp. 545-551  (1959).

\bibitem{Suzuki76}
M. Suzuki,
\textit{Generalized Trotter's formula and systematic approximants of exponential operators and inner
derivations with applications to many-body problems.} 
Commun. Math. Phys. {\bf 51,} 183--190 (1976).

\bibitem{Hatano-Suzuki}
N. Hatano and M. Suzuki,
\textit{Finding Exponential Product Formulas of Higher Orders.} 
arXiv:math-ph/0506007 (2005).

\bibitem{Leroux-2012}
I. D. Leroux, M. H. Schleier-Smith,  H. Zhang, and V. Vuleti\'{c},
\textit{Unitary cavity spin squeezing by quantum erasure.}
Phys. Rev. A  {\bf 85,} 013803 (2012).

\bibitem{Strobel-2014}
H. Strobel,   W. Muessel,   D. Linnemann,  T. Zibold, D. B. Hume,  L. Pezz\`{e},  A. Smerzi,  and M. K. Oberthaler,
\textit{Fisher information and entanglement of non-Gaussian spin states.}
Science  {\bf 345,} 424 (2014).

\end{thebibliography}
\end{document}